\documentclass{pnastwo}

\usepackage{graphicx}
\graphicspath{{Figures/}}

\url{www.pnas.org/cgi/doi/10.1073/pnas.0709640104}
\copyrightyear{2008}
\issuedate{Issue Date}
\volume{Volume}
\issuenumber{Issue Number}

\begin{document}

\title{Superradiance Transition and Nonphotochemical Quenching in Photosynthetic
Complexes}

\author{Gennady P. Berman\affil{1}{Theoretical Division, T-4, Los Alamos National Laboratory, and the New Mexico Consortium,  Los Alamos, NM 87544, USA},
Alexander I. Nesterov\affil{2}{Universidad de Guadalajara, Departamento de F{\'\i}sica, Jalisco, M\'exico}, 
Gustavo V. L\'opez  \affil{2}{}, \and
Richard T. Sayre\affil{3}{Biological Division, B-11, Los Alamos National Laboratory, and the New Mexico Consortium, Los Alamos, NM 87544, USA}}

\contributor{Submitted to Proceedings of the National Academy of Sciences
of the United States of America}

\date{\today}
\significancetext{GPB and AIN developed the concept of the study. GPB and RTS conducted the analysis, data interpretation, and drafted the manuscript. AIN contributed to the development of the numerical methods, data interpretation, and drafting of the manuscript. GLV contributed to the numerical simulations.}

\maketitle

\begin{article}
\begin{abstract}
{We demonstrate numerically that superradiance could play a significant role in nonphotochemical quenching (NPQ) in light-harvesting complexes. Our model consists of a network of five interconnected sites (discrete excitonic states) that are responsible for the NPQ mechanism. Damaging and charge transfer states are linked to their sinks (independent continuum  electron spectra), in which the chemical reactions occur. The superradiance transition in the charge transfer (or in the damaging) channel, occurs at particular electron transfer rates from the discrete to the continuum electron spectra, and can be  characterized by a segregation of the imaginary parts of the eigenvalues of the effective non-Hermitian Hamiltonian. All five excitonic sites interact with their protein environment that is modeled by a random stochastic process. We find the region of parameters in which the superradiance transition into the charge transfer channel takes place. We demonstrate that this superradiance transition  has the capability of producing optimal NPQ performance.}
\end{abstract}


\abbreviations{NQP, nonphotochemical quenching; RTP, random telegraph process; LHCs, light-harvesting complexes; ET, electron transfer.}

\dropcap{W}hen sunlight intensity is too high, damaging processes (such as singlet oxygen production) occur, that can destroy the photosynthetic organisms. To survive intense sunlight fluctuations, photosynthetic organisms have evolved many protective strategies, including  nonphotochemical quenching (NPQ) \cite{book1}. (See also references therein.) Using this strategy, light-harvesting complexes (LHCs) experience geometrical reorganizations due to the conformational changes of their protein environments. As a result, the damaging channels are partly suppressed, and the excessive sunlight energy is transfered to the non-damaging, quenching channel(s), that are opened in this regime.

In modeling the NPQ mechanisms, it is possible to characterize the  sites of the LHCs by the discrete excitonic energy states, $|n\rangle$,  $n$ being  the number of site, associated with the light-sensitive chlorophyll or carotenoid molecule. Then, both the damaging and undamaging channels can be characterized by their corresponding sinks, $|S_n\rangle$, that provide independent continuum electron energy spectra \cite{BNSS}. These sinks can have  very complex structures, and they can be responsible for primary charge separation processes, as in the photosynthetic reaction centers (RCs) \cite{Pudlak1,Pudlak2}, for creation of charge transfer states,  for  singlet oxygen production \cite{BNSS},  and for many other quasi-reversible chemical reactions.  Generally, a particular sink, $|S_n\rangle$, is connected to a particular site, $|n\rangle$. The energy transfer from this state to its sink is characterized by the electron transfer (ET) rate,  $\Gamma_n$. For those sites which are not connected to sinks, the corresponding ET rates vanish. Under reasonable assumptions, this type of  model can be described by an effective non-Hermitian Hamiltonian \cite{BNSS,Pudlak1,Pudlak2,Nes1,Nes2}. 

Then, this approach becomes similar to those used in describing the so-called ``superradiance transition" (ST) in systems in which the discrete (intrinsic) energy states interact with the continuum spectra \cite{Zel1,Ber2,Ber1,Zel2,Ber3,Ber4}. In these systems, the ST usually occurs when the resonances start to overlap. Namely, when the half-sum of the imaginary parts of the neighboring  eigenenergies (decay widths) becomes of the order of the spacing between the resonances (distance between the real parts of the neighboring eigenenergies). 

For small ET rates, the decay widths are small, and resonances are isolated. However, if the $\Gamma_n$  approach  critical values, at which the resonances  begin to overlap, then a segregation of the widths occurs. In this case, most of the decay width is allocated to  a few short-lived ``superradiant" states, while all other states become long-lived (and effectively decoupled from the environment). This segregation is called the ST.  The ST is a quantum coherent effect and, in biological systems, it should be considered (see, for example, \cite{Ber3,Ber4}) in close relation to the quantum coherent effects studied recently in the photosynthetic complexes \cite{QEB}.

It is important to note, that in many systems the occurrence of the ST  requires not only the overlapping of the neighboring resonances, but also a delocalization of the eigenfunctions which are involved in the ST. This is especially important for those systems in which the initial wave function does not sufficiently overlap with those eigenfunctions which are responsible for the SR  (see below). 

The ST has been well-studied in random matrix theory \cite{Ber2,Ber1,VWZ,LSSS,FS,SFT},  nuclear physics \cite{VZ1,VZ2,VZ3},  microwave billiards \cite{SP,NS}, and in paradigmatic models of coherent quantum transport \cite{CK,CSSZ,SR}.  In particular, it was shown in \cite{CK,CSSZ} that, for a realistic model for quantum transport,  maximum transmission is achieved at the ST. In biological systems, the ST has previously been discussed in \cite{Ber3,Ber4,CB1,PW}. (See also references therein.)

In this paper, we numerically study the possible role of the ST in the NPQ mechanism in the LHCs of photosystem II (PSII). For concreteness, we apply our approach to the NPQ protocol discussed in \cite{Ahn}. According to \cite{Ahn}, the NPQ mechanism is associated with charge transfer quenching. It takes place in part in the LHC CP29 minor LHC,  which is located (with other minor LHCs) between the RC and the peripheral LHCII complex \cite{Bennett}. In the NPQ mechanism, discussed in \cite{Ahn}, the following basic components are involved: a dimer based on two chlorophylls with their excited states, $Chla_{5}^*$ and $Chlb_{5}^*$; the heterodimer, $(Chla_{5}-Zea)^*$, where the carotenoid zeaxanthin, $Zea$, is formed in the NPQ regime; the charge transfer state (CTS) of this heterodimer,  $(Chla_{5}^--Zea^+)^*$; and the damaging channel, which is responsible for production of the singlet oxygen and other damaging products. Our model (see also \cite{BNSS}) includes five discrete energy levels (sites), each for a mentioned above component. Each of two discrete levels, associated with the CTS and with the damaging channel, is connected to its corresponding independent sink, which is described by the continuum electron spectrum. Both these sinks are responsible for the quasi-reversible chemical reactions. According to \cite{Ahn}, in the NPQ regime, the extra sunlight energy, that is absorbed by LHC, is transfered to the CTS, and then it partly dissipates into the protein environment. 

In our model, all five discrete states and two sinks are embedded in the protein environment, which we model by a random telegraph process (RTP). In \cite{BNSS}, our model was used to simulate numerically the efficiency of the NPQ regime, at various parameters of the system. In the present paper, we extend the application of our model to describe the possible role of the ST in the NPQ regime. Our mathematical approach is based on a non-Hermitian Hamiltonian which naturally occurs and is a standard mathematical tool used for studying the ST in many physical and biological systems. We demonstrate that, for reasonable conditions, the NPQ mechanism can be associated with the ST in the charge transfer channel. We also show that the ST can optimize the performance of the NPQ mechanism. 

\section{Mathematical formulation of the model}

\begin{figure}[ht]
	\begin{center}
		\scalebox{0.325}{\includegraphics{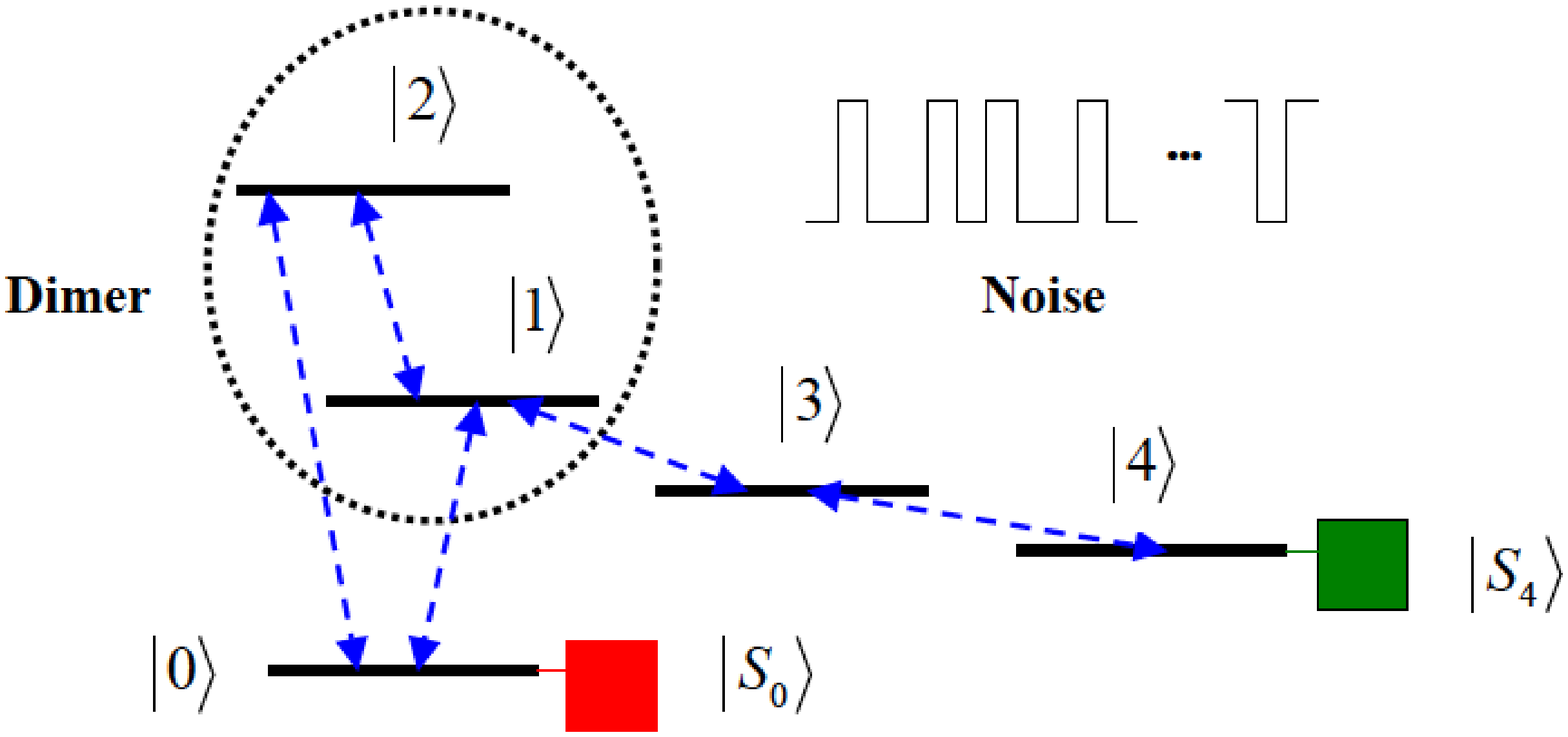}}
	\end{center}
	\caption{Schematic of our  model consisting of five discrete  states, $|n\rangle$, $(n=0,...,4)$, and two independent sink reservoirs, $|S_0 \rangle$ (connected to the damaging state), and  $|S_4 \rangle$ (connected to the charge transfer state). The ellipse indicates the dimer based on excited states of two chlorophylls. The blue dashed lines indicate non-zero matrix elements used in numerical simulations. 
		\label{M}}
\end{figure}

  Our model includes five discrete excitonic energy states  and two sinks  (see Fig. \ref{M}). Below, we use the notation and the basic NPQ components similar to those used in \cite{BNSS,Ahn}.  The values of the parameters are varied in order to demonstrate the different dynamical regimes of the NPQ mechanism and their relation to the ST. We use the following notation. The discrete  electron state, $|1\rangle\equiv |Chla_5^*\rangle$, is the excited electron state of $Chla_5$.  The discrete  electron state, $|2\rangle\equiv |Chlb_5^*\rangle$, is the excited electron state of $Chlb_5$.  So, in our model a dimer based on two chlorophylls, $Chla_5$ and $Chlb_5$, each in an excited electron state, participates in the NPQ process, as in \cite{Ahn}. The discrete state,  $|3\rangle\equiv |(Chla_5-Zea)^*\rangle$, is the heterodimer excited state.  The discrete state,  $|4\rangle\equiv |(a^{-}-Zea^{+})^*\rangle$, is the CTS of the heterodimer. The state, $|0\rangle\equiv |damage\rangle$, is a discrete part of the damaging  channel. The sink, $|S_0\rangle$, is the continuum part of the damaging channel. The sink, $|S_4\rangle$, is the continuum part of the CTS (channel), $|4\rangle$. All matrix elements, $V_{mn}$, in the Hamiltonian (\ref{Ham}) describe the interactions between the discrete excitonic states. Each sink is characterized by two parameters: the rate, $\Gamma_{0}$ ($\Gamma_{4}$), of  ET into the sink, $|S_{0}\rangle$ ($|S_{4}\rangle$), from the corresponding attached discrete state, $|0\rangle$ ($|4\rangle$), and the efficiency (cumulative time-dependent probability), $\eta_{0}(t)$ ($\eta_{4}(t)$), for the exciton to be absorbed by the corresponding sink.  Note that both $\Gamma_0$ and $\Gamma_4$ characterize only the rates of destruction of the exciton in the CP29 LHC, and they do not describe any subsequent chemical reactions that take place in the damaging and the charge transfer channels. In this sense, our model describes only the primary NPQ processes in the ET, and it does not describe the processes which occur in both sinks. The latter occur on relatively large time-scales, and require a detailed knowledge of the structures of the sinks, and additional methods for their analysis.

According to \cite{BNSS}, the quantum dynamics of the ET can be described by the following effective non-Hermitian Hamiltonian,
 $ \tilde{\mathcal H}= {\mathcal H}- i \mathcal W$, where,
 \begin{eqnarray}\label{Ham}
 {\mathcal H} = \sum_{n}\varepsilon_n |n\rangle \langle n|  +\sum_{m\neq n} V_{mn} |m\rangle \langle n|, \quad m,n =0,1,\dots, 4,
 \end{eqnarray}
is the dressed Hamiltonian, and
 \begin{eqnarray}
 \label{Gamma}
{\mathcal W} = \frac{1}{2}\sum_{n=0,4}\Gamma_n|n\rangle \langle n|.
 \end{eqnarray}

 In (\ref{Ham}), $\varepsilon_n$ is the renormalized energy of the state, $|n\rangle$,	
 	and the parameter, $\Gamma_n$, in (\ref{Gamma})  is the tunneling rate to the $n$-{th} sink.  Below, only $\Gamma_0$ and $\Gamma_4$ are taken into account.

The dynamics of the system can be described by the Liouville-von Neumann equation,  \begin{eqnarray}\label{DM1}
    \dot{ \rho} = i[\rho,\mathcal H] - \{\mathcal W,\rho\},
 \end{eqnarray}
 where $\{\mathcal W,\rho\}= \mathcal W\rho +\rho\mathcal  W$.

We define the ET efficiency of tunneling to all $N$ sinks as,
\begin{eqnarray}\label{ET1} 
\eta(t) = 1 - {\rm Tr}(\rho(t)) =  \int_0^t {\rm Tr}\{\mathcal W,\rho(\tau)\} d \tau.
\end{eqnarray}
This can be expressed as the sum of time-integrated probabilities of trapping an electron into the $n$-th sink, $\eta(t)=\sum_{n}\eta_n(t)$ \cite{Lloyd1,CDCH}. Here, 
\begin{eqnarray}
\eta_n(t) = \Gamma_n \int_0^t \rho_{nn}(\tau)d \tau,
\label{Eq16ar}
\end{eqnarray}
is the efficiency of the $n$-th sink. In particular, for our model, presented in Fig. \ref{M}, complete suppression of the damaging channel {\it at all times} occurs if  $\eta_0(t)=0$.

\subsection{The eigenvalue problem for the non-Hermitian Hamiltonian}

The bi-orthogonal eigenstates of the effective non-Hermitian Hamiltonian, $\tilde{\mathcal H}$, provide a convenient basis in which the  eigenvalue problem can be formulated and resolved \cite{Zel1,Zel2}. (See also references therein.) Let $|\psi_\alpha \rangle$ be the right eigenstates of $\tilde{\mathcal H}$ with complex eigenvalues, ${\tilde E}_\alpha$,
\begin{eqnarray}\label{EP}
	\tilde{\mathcal H}|\psi_\alpha\rangle = {\tilde E}_\alpha |\psi_\alpha \rangle ,\quad \alpha =1,2 \dots 5 \,.
\end{eqnarray}
Further it is convenient to set ${\tilde E}_\alpha = {\mathcal E}_\alpha - i \Upsilon_\alpha $, where, ${\mathcal E}_\alpha$ and $\Upsilon_{\alpha}$ (decay width), are the real and the negative imaginary parts of the complex eigenvalue.

By manipulating the tunneling rates into the sinks, one can modify the dynamics of the whole system. As will be demonstrated below, under some conditions, quantum coherent escape of the exciton in the charge transfer channel  occurs. This escape is associated with the ST in the sink, $|S_4\rangle$. A similar ST can be realized, under some conditions, in the damaging channel (sink $|S_0\rangle$).

\subsection{Noise-assisted ET in the NPQ regime}

 In the presence of noisy environment, the evolution of the system can be described by the following effective non-Hermitian Hamiltonian \cite{BNSS}, 
 \begin{eqnarray}
 \tilde{\mathcal H}_{tot}= {\mathcal H}- i \mathcal W + {\mathcal V}(t),
 \end{eqnarray}
 where, ${\mathcal V}(t)=  \sum_{m,n} \lambda_{mn}(t)|m\rangle\langle  n |
 , \quad m,n = 0,1,\dots 4$, $\lambda_{mn}(t)$, describes the influence of the protein noisy environment.  The diagonal matrix elements of noise, $\lambda_{nn}(t)$, are responsible for decoherence. The off-diagonal matrix elements, $\lambda_{mn}(t)$, lead to {\it direct} relaxation processes. In what follows, we restrict ourselves to the diagonal noise effects. (See also \cite{Marcus1,Xu,Ber5}, and references therein.) Then, one can write, $\lambda_{mn}(t) = \lambda_n \delta_{mn}\xi(t) $,  where $\lambda_n$ is the coupling constant at site, $n$, and $\xi(t)$ is a random process.

 We describe the protein noisy environment by an RTP,  $\xi(t) $, with the following properties,
 \begin{eqnarray}\label{chi_8}
&\langle \xi(t)\rangle =0, \\
 &\chi(t-t')=\langle \xi(t)\xi(t')\rangle,
\end{eqnarray}
where, $\chi(t-t') = \sigma^2 e^{-2\gamma |t-t'|}$, is the correlation function of noise. 

 The evolution of the  average  components of the density matrix is described by the following system of ordinary differential equations \cite{BNSS}:
\begin{eqnarray} \label{IB4}
&&\frac{d}{dt}{\langle{\rho}}\rangle =i[\langle\rho\rangle,\mathcal H] - \{\mathcal W,\langle\rho\rangle\} - i B\langle\rho^\xi\rangle, \\
&&\frac{d}{dt}{\langle{\rho^\xi}}\rangle =i[\langle\rho^\xi\rangle,\mathcal H] - \{\mathcal W,\langle\rho^\xi\rangle - i B\langle\rho\rangle- 2\gamma \langle\rho^\xi\rangle ,
\label{IB5}
\end{eqnarray}
where $\langle\rho^\xi\rangle = \langle\xi\rho \rangle/\sigma$, $B = \sum_{m,n}(d_m - d_n)|m\rangle \langle n|$, and we set $d_{m}=\lambda_m \sigma$. The average,  $\langle \,\dots \,\rangle$, is taken over the random process. In deriving Eqs. (\ref{IB4}) and (\ref{IB5}), the approach developed in \cite{KV1,KV2,KV3} for the RTP was used.

Employing Eqs. (\ref{IB4}) and \ref{IB5}, one can show that  the following normalization condition is satisfied,
 \begin{equation}
 	\label{C1}
 	\sum_{n=0}^4 \langle\rho_{nn}(t)\rangle+\eta_0(t)+\eta_4(t)=1.
 \end{equation}
Eq. (\ref{C1})  requires that the total probability of finding the exciton among the five discrete levels and in two sinks is unity for all times. 

\section{Results of numerical simulations}

Here we examine numerically the possible role of the ST in the NPQ regime. The results of our numerical simulations are obtained by solving Eqs. (\ref{IB4}) and (\ref{IB5}) for the average components of the density matrix. Below, in numerical simulations, we choose $\hbar= 1$. Then, the values of parameters in energy units are measured in $\rm ps^{-1}$ ($1\rm ps^{-1} \approx 0.66\rm {meV}$). Time is measured in $\rm ps$.
 
To simplify the presentation of our results, some characteristic parameters were fixed. The following values of these parameters were chosen:  The energy of the excited state of $Chla_5^*$, $\varepsilon_1=60\rm ps^{-1}\approx 39.6\rm meV$;  The energy of the excited state of $Chlb_5^*$, $\varepsilon_2=210\rm ps^{-1}\approx 138.6\rm meV$; The energy of the excited state of the heterodimer, $(Chla_5-Zea)^*$, $\varepsilon_3=45\rm ps^{-1}\approx 29.7\rm meV$; The energy of the CTS  of the heterodimer, $(Chla_5^{-}-Zea^{+})^*$, $\varepsilon_4=30\rm ps^{-1}\approx 19.8\rm meV$. The following values of the matrix elements of the dipole-dipole interactions between the sites were fixed: $V_{10}=V_{20}=30\rm ps^{-1}\approx 19.8\rm meV$, $V_{12}=15\rm ps^{-1}\approx 9.9\rm meV$, $V_{13} = V_{34}=25\rm ps^{-1}\approx 16.5\rm meV$. The values of the coupling constants were chosen: 
 $d_1=\varepsilon_1$, $d_2=\varepsilon_2$, $d_3=\varepsilon_3$, $d_4=\varepsilon_4$. These values of the coupling constants provide the strongest (resonant) noise on the corresponding sites (see below). The correlation decay rate was fixed: $\gamma=10\rm ps^{-1}$. The time duration for simulation of the ET efficiencies, $\eta_0(\tau) $ and $\eta_4(\tau)$, was fixed at $\tau=5\rm ps$.  The parameters:  $\varepsilon_0$ (characterizing the energy level of the damaging state, $|0\rangle$); the coupling constant, $d_0$, of the state, $|0\rangle$, with noise; and the ET rates, $\Gamma_0$ and $\Gamma_4$, were varied in some simulations. The initially populated state was chosen to be the excited state of $Chla_5^*$.
 
\begin{figure*}[ht]
 	\begin{center}
 		\scalebox{0.425}{\includegraphics{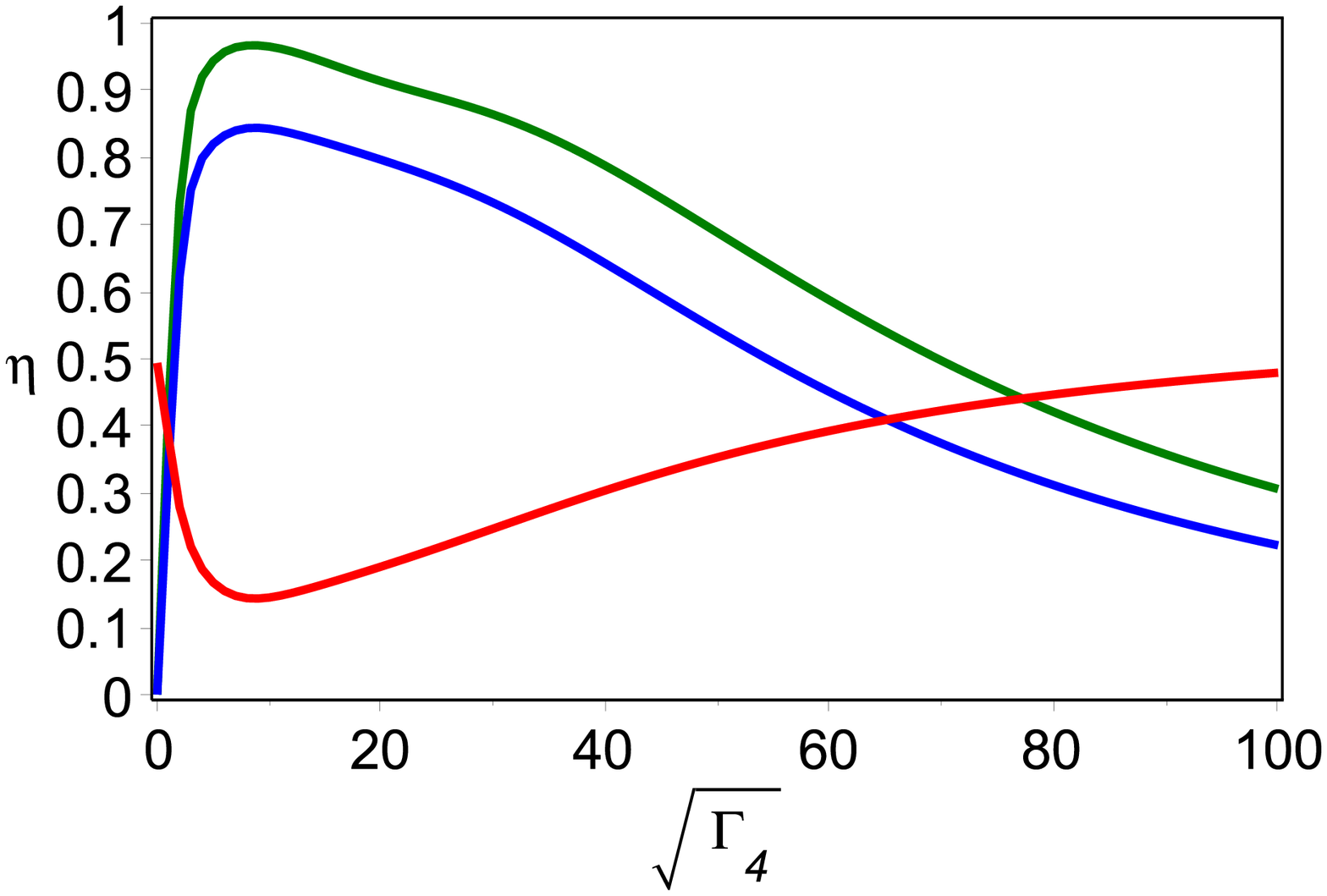}}	
 		(a)
 		\scalebox{0.4}{\includegraphics{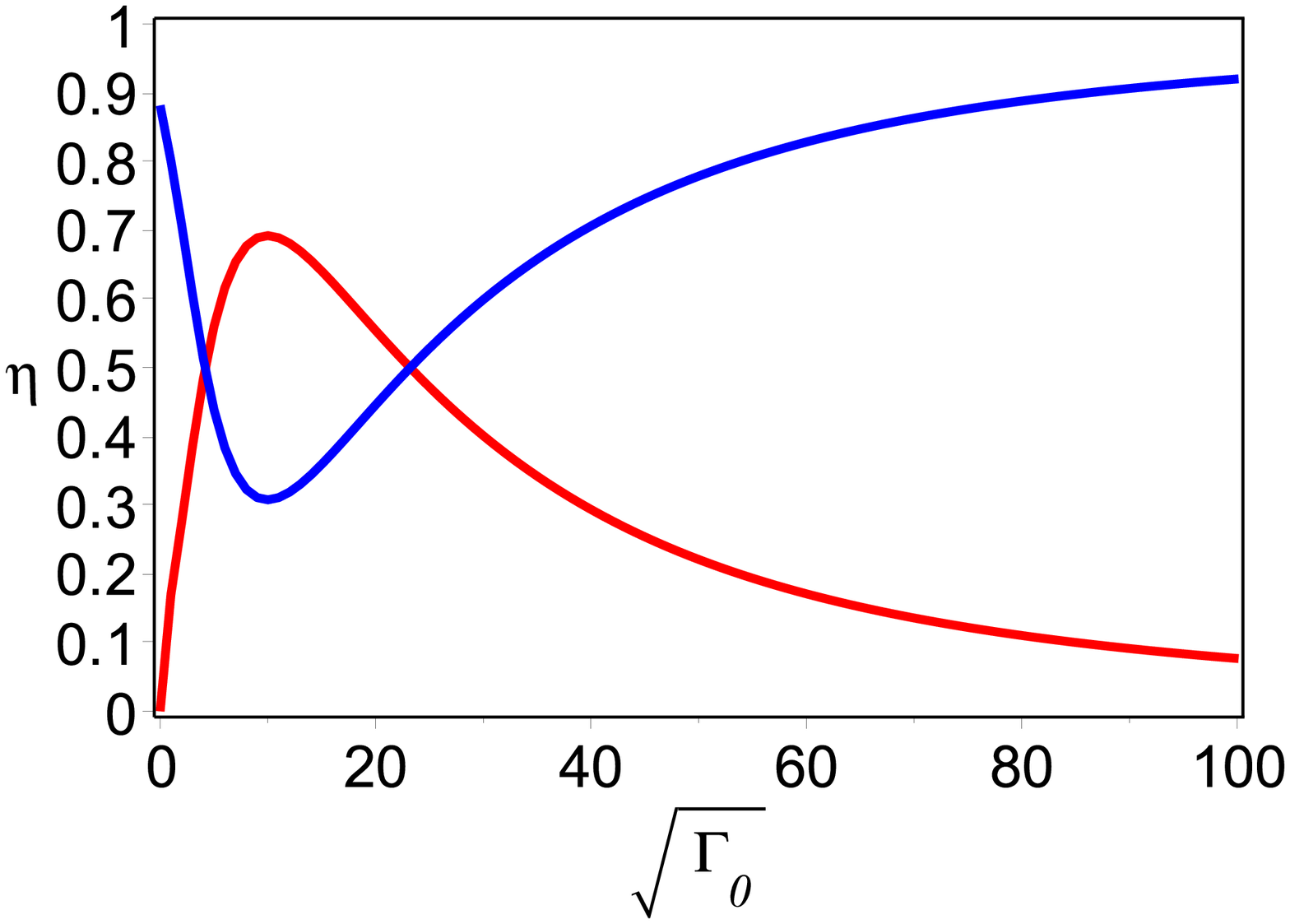}}
 		(b)
 	\end{center}
 	\caption{(Color online) The efficiencies of two sinks in the absence of noise ( $\varepsilon_0=0$). (a) The efficiencies of sinks, $\eta_4$ (blue and green curves) and $\eta_0$ (red curve)  vs $\sqrt{\Gamma_4}$, for  $\Gamma_0 =0$ (green curve) and for $\Gamma_0 =2$ (blue and red curves); (b) The efficiencies of sinks,  $\eta_4$ (blue curve) and $\eta_0$ (red curve) vs. $\sqrt{\Gamma_0}$, for $\Gamma_4 =10$.
 		\label{A1}}
 \end{figure*}

 In Figs. \ref{A1}a and \ref{A1}b, we computed the efficiencies of both sinks, $\eta_{0,4}(\tau)$, in the absence of noise. The energy of the damaging state, $|0\rangle$, was chosen: $\varepsilon_0=0$, close enough to initially populated level, $\varepsilon_1$, of $Chla_5^*$.  
 As one can see from Fig. \ref{A1}a, when only one sink, $|S_4\rangle$ (related to the CTS) is open (green curve, $\Gamma_0=0$), the efficiency of the NPQ process experiences a maximum ($\eta_4^{max}\approx 97\%$), at the ET rate, $\Gamma_4^{0}\approx 72\rm ps^{-1}$. The important observation is that when $\Gamma_4$  increases, the efficiency, $\eta_4$, decreases.  As will be discussed below, this  dependence of the efficiency, $\eta_4$ on $\Gamma_4$ is caused by the ST. When two sinks are open ($\Gamma_0=2\rm ps^{-1}$), the efficiency of the NPQ process (blue curve in Fig. \ref{A1}a) still has a maximum as a function of the ET rate, $\Gamma_4$. In this case, $\eta_4^{max} \approx 82\%$ and $\Gamma_4^{0}\approx 72\rm ps^{-1}$. This decrease of the NPQ efficiency is caused by the damaging channel, $|S_0\rangle$, during $\tau=5\rm ps$ (red curve in Fig. \ref{A1}a).  In Fig. \ref{A1}b, the opposite case is considered, when the ET rate to the CTS (sink $|S_4\rangle$) is fixed ($\Gamma_4=10\rm ps^{-1}$), and both efficiencies, $\eta_0$ and $\eta_4$, are simulated as functions of the ET rate,  $\Gamma_0$, to the damaging sink. Similar to the results presented in Fig. \ref{A1}a,  in Fig. \ref{A1}b one can observe the ST in the damaging sink, $|S_0\rangle$. In this case, the maximal efficiency is relatively small, $\eta_0^{max}\approx 70\%$, because the ET rate to the CTS is relatively high ($\Gamma_4=10\rm ps^{-1}$).
 
\begin{figure*}[ht]
 	\begin{center}
 		\scalebox{0.4}{\includegraphics{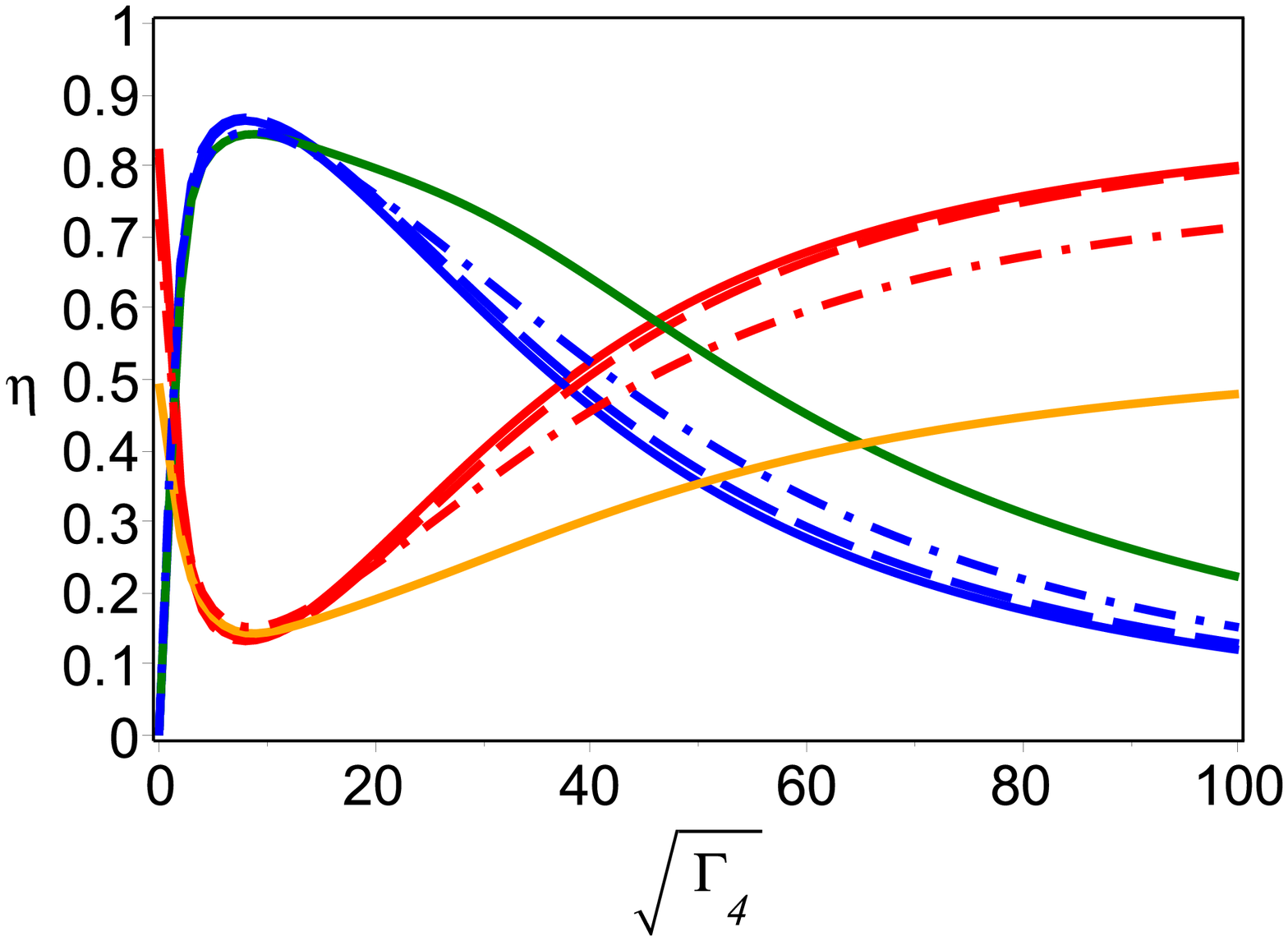}}	
 		(a)
 		\scalebox{0.415}{\includegraphics{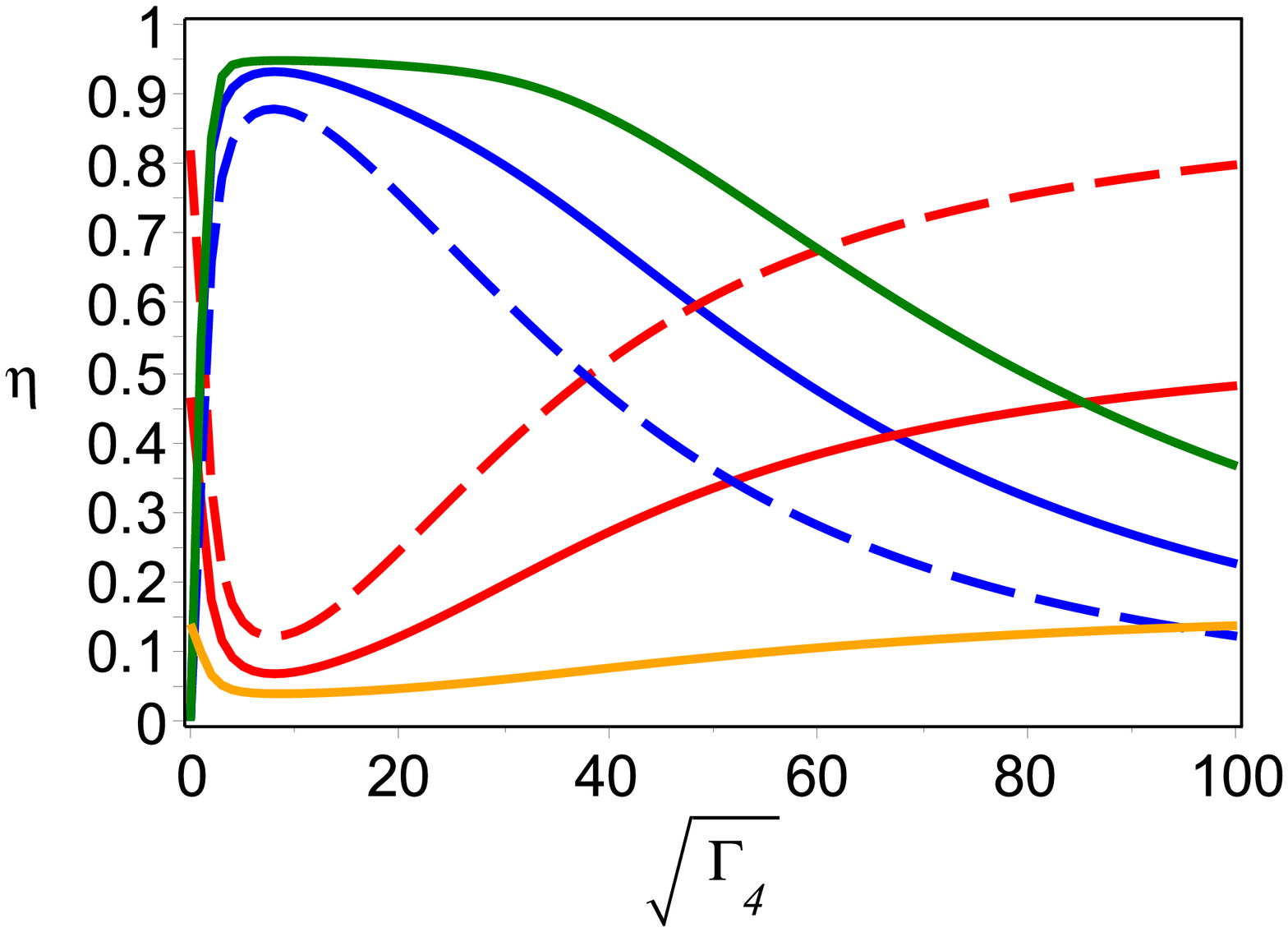}}	
 		(b)
 	\end{center}
 	\caption{(Color online)  The efficiencies of two sinks in the presence of noise: $\eta_4$ (blue curve) and $\eta_0$ (red curve)  vs. $\sqrt{\Gamma_4}$. (a) $~\varepsilon_0=0$: solid curves: resonance noise for dimer $1-0$  ($d_0=0$); dashed curves: $d_0=-30$; dash-dot curves: $d_0=30$. (b) $~\varepsilon_0=-90$: solid curves ($d_0=0$); resonance noise, $d_0=-90$ (dashed curves). $\Gamma_0 =2$.  Green and orange curves demonstrate  the efficiencies of sinks in the absence of noise.	
 		\label{A3}}
 \end{figure*}

 In Figs. \ref{A3}a and \ref{A3}b, the results of numerical simulations are presented for the NPQ efficiency, $\eta_4(\Gamma_4)$ and  for the efficiency of the damaging channel, $\eta_0(\Gamma_4)$, for fixed ET rate, $\Gamma_0=2\rm ps^{-1}$. So, two sinks are open in this case. The two main differences from the previous results presented in Figs. \ref{A1}a and \ref{A1}b, are caused by the following two modifications. First, in Fig. \ref{A3}b, the energy of the damaging state was chosen, $\varepsilon_0=-90\rm ps^{-1}\approx -59.4\rm meV$, which is significantly below the value, $\varepsilon=0$, which was used in Figs. \ref{A1}a and \ref{A1}b (and in Fig. \ref{A3}a). Second, the influence of noise, produced by the protein environment, is included. 
 
 In our model, noise is generated by the RTP, and it has two characteristic parameters: the amplitude of the correlation function, $\sigma^2$, and the correlation time, $\tau_c=2/\gamma$. Also,
 the influence of noise on the CP29 LHC is characterized by the interaction constants (matrix elements of ``site-protein" interactions), $\lambda_n$ ($n=0,1,..,4$). Then, instead of parameters, $\lambda_n$, the renormalized parameters, $d_n=\lambda_n\sigma$, occur in our model. The important property of the influence of noise on the chlorophyll dimer, discussed in \cite{Ber6,NB5}, is the following. Under the ``resonant conditions", $\varepsilon_n-\varepsilon_m=d_n-d_m$, the influence of noise on the dimer, $n-m$, becomes maximal in the sense that the ET rate has a resonant peak.
 
 In Fig. \ref{A3}a, the damaging level ($\varepsilon_0=0$) is chosen close to the energy level of $Chla_5^*$.  The efficiencies of sinks, $\eta_4$ and $\eta_0$, are indicated by the blue and red curves, respectively. Blue  and red solid curves  correspond to the resonant noise ($d_0=0$) on the dimer, $1-0$. Blue and red dashed and blue and red dash-dot curves correspond to non-resonant noise on the dimer, $1-0$. Green and yellow curves correspond to the absence of noise. As one can see, the maximum of the NPQ efficiency is, $\eta_4^{max}\approx 85\%$, almost independent of noise, in the vicinity of $\Gamma_4^{0}\approx 72\rm ps^{-1}$. This behavior of $\eta_4$, in the vicinity of  $\Gamma_4^{0}$, can be explained as follows. Because the damaging sink is close (in energy) to the $Chla_5^*$ energy level, the damaging sink accumulates during $\tau=5\rm ps$ the amount of probability which practically is independent of the amplitude of noise.
 
 Different dependence of the NPQ efficiency on noise, in the vicinity of its maximum, occurs when  the damaging energy level, $\varepsilon_0=-90\rm ps^{-1}\approx 59.4\rm meV$, is relatively far from the energy level of the $Chla_5^*$. As one can see from  Fig. \ref{A3}b, the resonant noise ($d_0=\varepsilon_0$), acting on the dimer, $1-0$, significantly degrades  the NPQ performance in the vicinity of $\eta_4^{max}\approx 88\%$ (blue dashed curve). The reason is that the resonant noise acting on the $1-0$ transition, helps to accumulate extra probability in the damaging sink. When noise is non-resonant on the $1-0$ transition (blue curve) the NPQ efficiency improves up to, $\eta_4^{max}\approx 95\%$, and it approaches the NPQ efficiency when noise is absent (green curve).
 
 \subsection{The superradiance transition in relation to the NPQ}
 
\begin{figure*}[ht]
	\begin{center}
		\scalebox{0.4}{\includegraphics{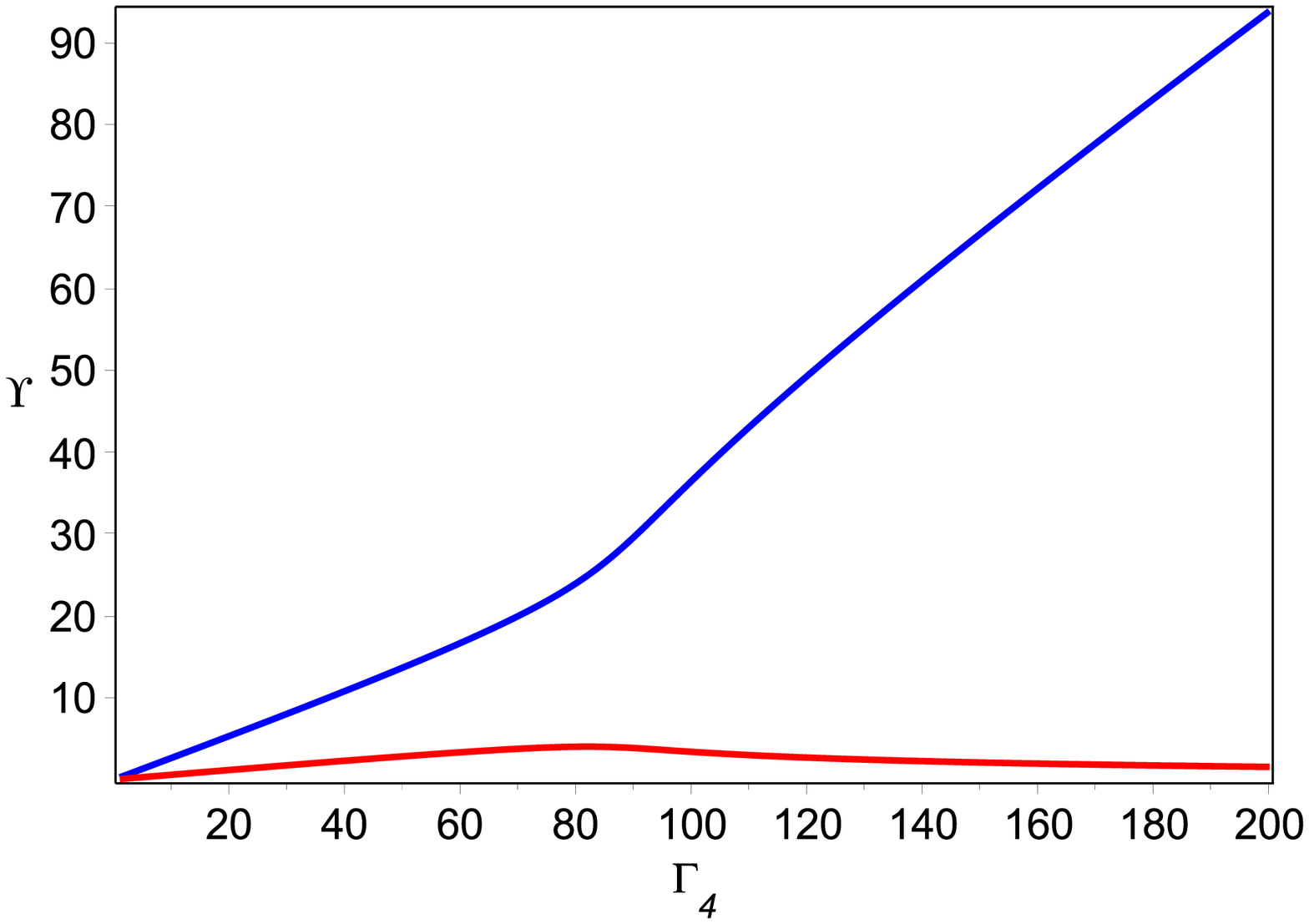}}	
		(a)
		\scalebox{0.35}{\includegraphics{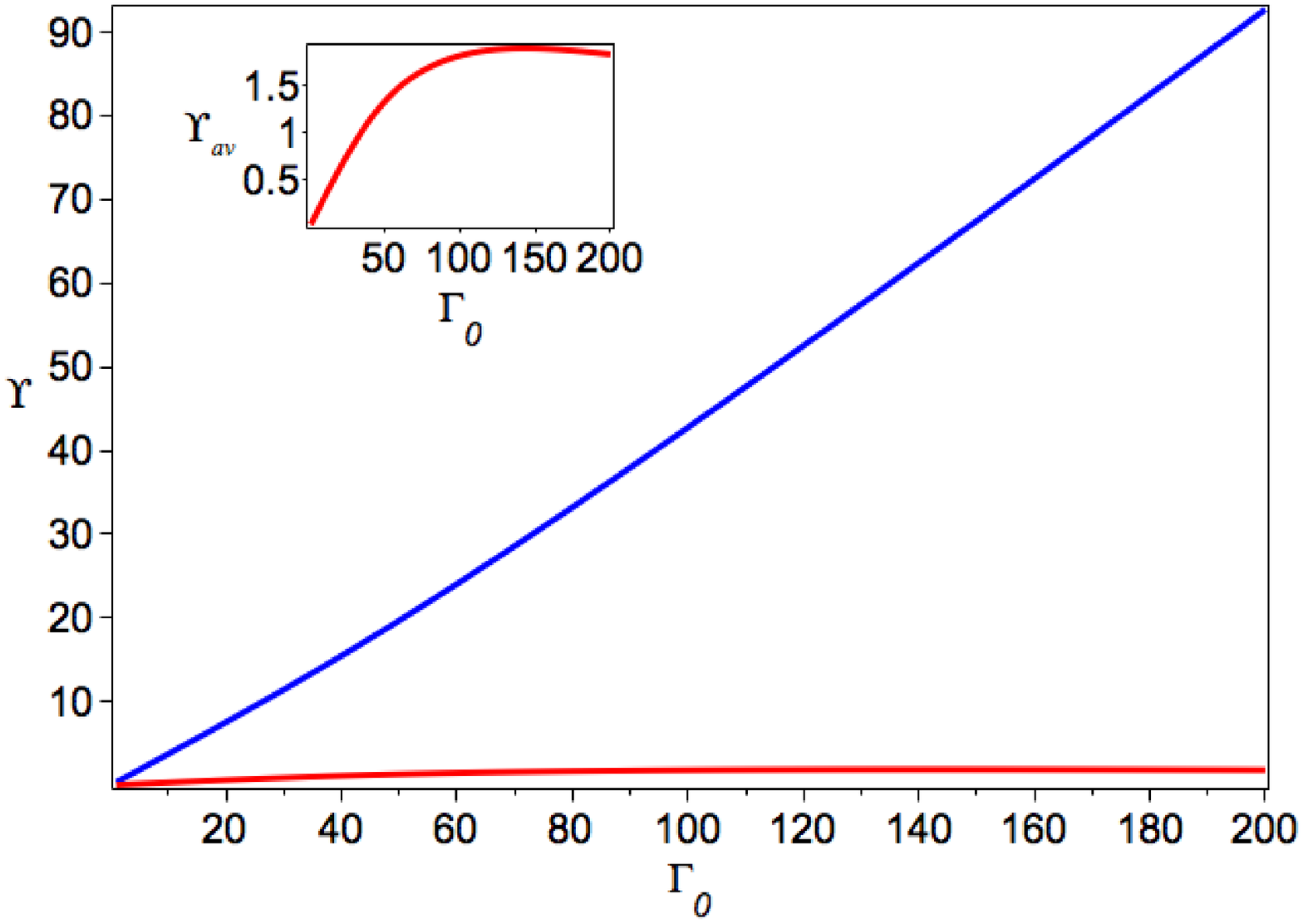}}	
		(b)
			\end{center}
	\caption{(Color online) Decay width of eigenstate as a function of the rate: (a) $\Gamma_4$ ( $\Gamma_0 = 0$);  (b) $\Gamma_0$ ( $\Gamma_4 = 0$).  The blue curve shows the largest decay width. The red curve represents the average decay width, ${\Upsilon}_{av}$, of the 4 states with smallest widths. Inset: zoom of ${\Upsilon}_{av}$. ($\varepsilon_0=-90$).
	\label{G2}}
\end{figure*}

 When noise is absent, the ST, in relation to the NPQ mechanism, is easy to explain in terms of the effective non-Hermitian Hamiltonian,  $\tilde{\mathcal H}= {\mathcal H}- i \mathcal W$, where the Hermitian operator, ${\mathcal H}$, is defined by Eq. (\ref{Ham}), and the Hermitian operator, $\mathcal W$, is defined by Eq. (\ref{Gamma}).  In Eq. (\ref{Gamma}), only $\Gamma_0$ and $\Gamma_4$ differ from zero. The eigenvalue problem for $\tilde{\mathcal H}$ is formulated in Eq. (\ref{EP}).
 In Fig. \ref{G2}a, the blue curve corresponds to the negative imaginary part (decay width) of the eigenvalue, $\Upsilon_\alpha$, with $\Upsilon_{max}$, as a function of $\Gamma_4$ (blue curve). The red curve corresponds to the average decay width, $\Upsilon_{av}$, of 4 eigenstates with smallest values of $\Upsilon_\alpha$, as a function of $\Gamma_4$. The energy level of the damaging state was chosen, $\varepsilon_0=-90\rm ps^{-1}\approx -59.4\rm meV$. As one can see, in the vicinity of $\Gamma_4^{*}\approx 84\rm ps^{-1}$, the red curve experiences a maximum. At this point, a segregation of the decay widths begins. Namely, when $\Gamma_4$ grows, the red and blue curves start to separate. So, the decay width of one (superradiant) eigenstate continue to grow with $\Gamma_4$, and the decay widths of all other (subradient) eigenstates decreases. This segregation behavior of the decay widths is one of the characteristic properties of the ST \cite{Zel1,Zel2}. As one can see from Fig. \ref{A3}b (green curve), in the vicinity of the ST ($\Gamma_4^{*}\approx 84\rm ps^{-1}$), the NPQ efficiency experiences its maximum behavior, with $\eta_4^{max}\approx 95\%$.
 
 Similar ST behavior occurs in Fig. \ref{G2}b, when $\Gamma_4=0$, and the dependence of widths on tunneling rate into the sink $|S_0 \rangle$ is presented. The insert shows the detailed behavior of the red curve. In this case, the ST occurs in the damaging sink, $|S_0\rangle$.

 \subsection{Overlapping of resonances} 
\begin{figure*}[ht]
\begin{center}
		\scalebox{0.4}{\includegraphics{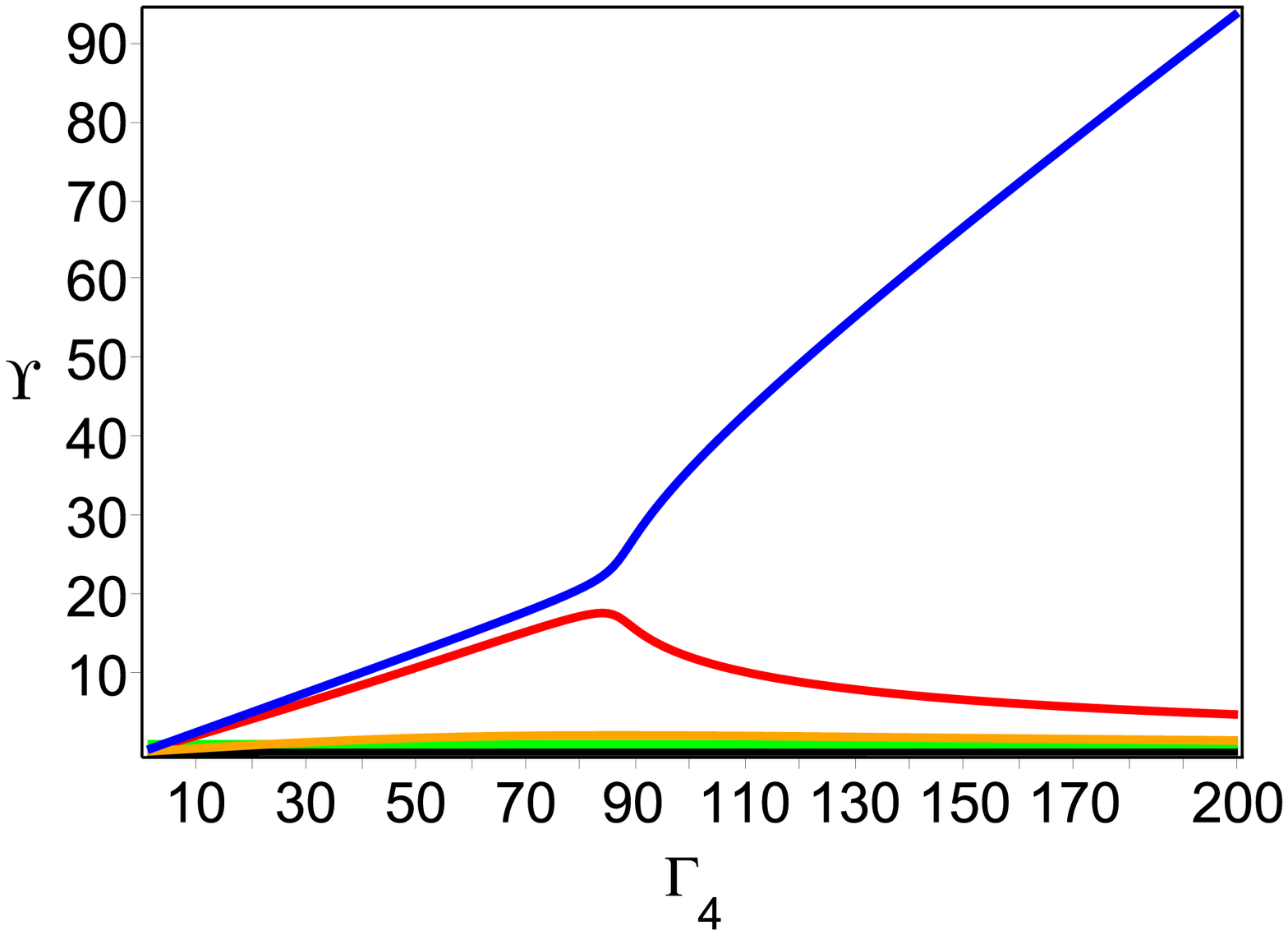}}	
		(a)
		\scalebox{0.4}{\includegraphics{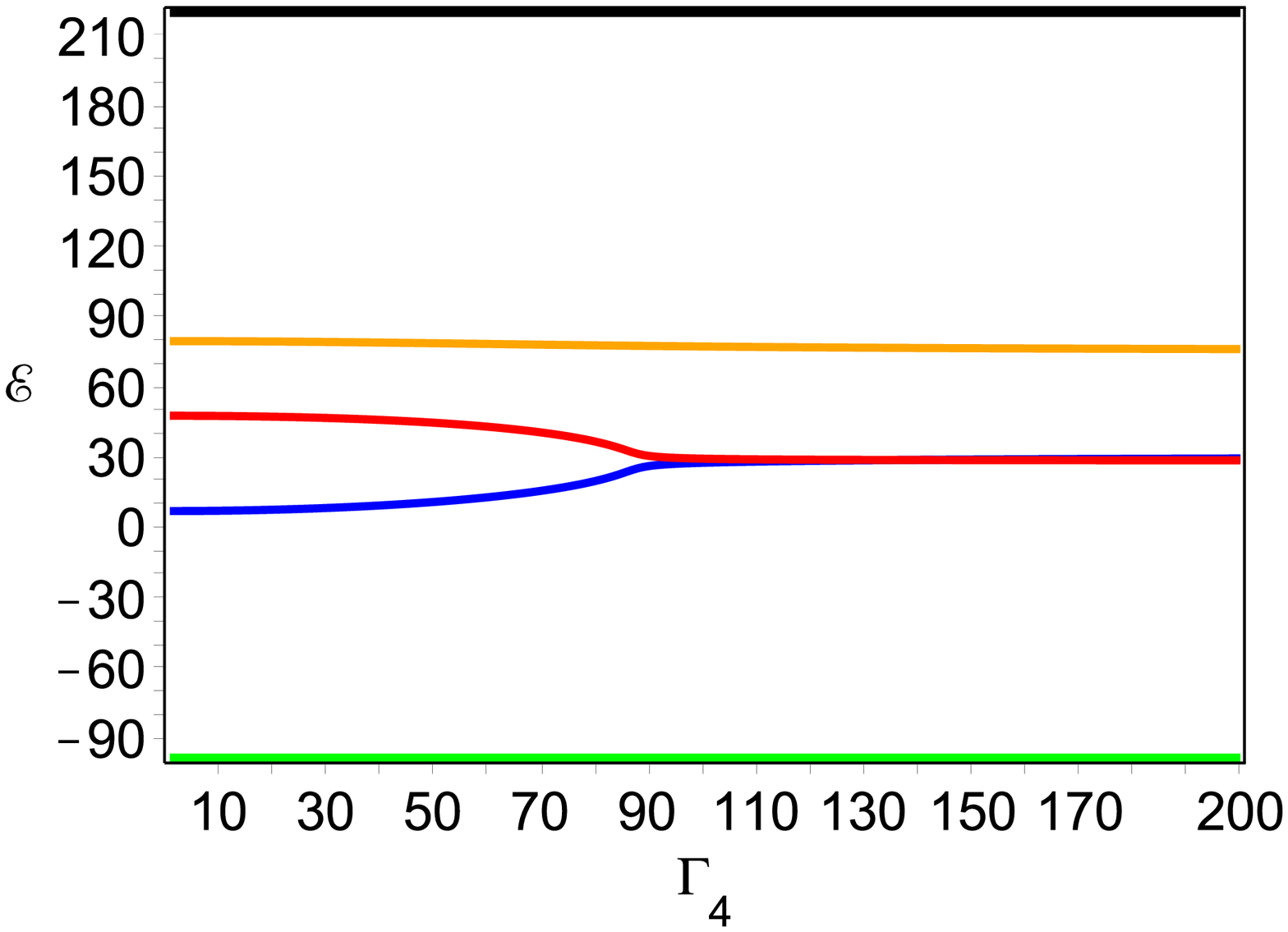}}	
(b)
			\end{center}
		\caption{(Color online) (a) Decay widths as  functions of the rate $\Gamma_4$.   (b) ${\mathcal E} =\Re(\tilde E)$ vs ${\Gamma_4}$. The blue curves correspond to the largest decay width.  ($\varepsilon_0=-90,~\Gamma_0=2$).
		\label{D1}}
\end{figure*}

In Figs. \ref{D1}a and \ref{D1}b, we show the dependences on $\Gamma_4$ of the decay widths, $\Upsilon_\alpha$, and the real parts of the eigenvalues, ${\mathcal E_\alpha}$. Different colors correspond to five different eigenstates. As one can see, at the superradiance transition ($\Gamma_4^*\approx 84\rm ps^{-1}ps$), the superradiant state, with largest decay width (blue curve), begins to overlap (interact) with its neighboring eigenstate (red curve). Indeed, at this point the spacing between these two states (resonances) is approximately equal to the half-sum of their decay widths: $\mathcal E_{blue}-\mathcal E_{red}\approx \Upsilon_{blue}-\Upsilon_{red}$. Then, these resonances cannot be considered to be isolated, and they both  coherently contribute to the ET to the corresponding sink (continuum) \cite{Zel1,Zel2}. In our case, only two resonances overlap at $\Gamma_4^*$. When $\Gamma_4$ grows, the decay width of the superradiant state grows, and finally the superradiant resonance overlaps with all other resonances. (See Figs. \ref{D1}a and \ref{D1}b.) But this does not mean that the superradiant state will continue to contribute (for large ET rates, $\Gamma_4$) to the ET into the continuum (sink). The localization properties of the eigenstates become important. (See the next sub-section.) Note also, that usually many interacting resonances may overlap and contribute to the ST \cite{Zel1,Zel2}.
 
 \subsection{The trajectories of the eigenenergies}
 
\begin{figure*}[ht]
	\begin{center}
		\scalebox{0.4}{\includegraphics{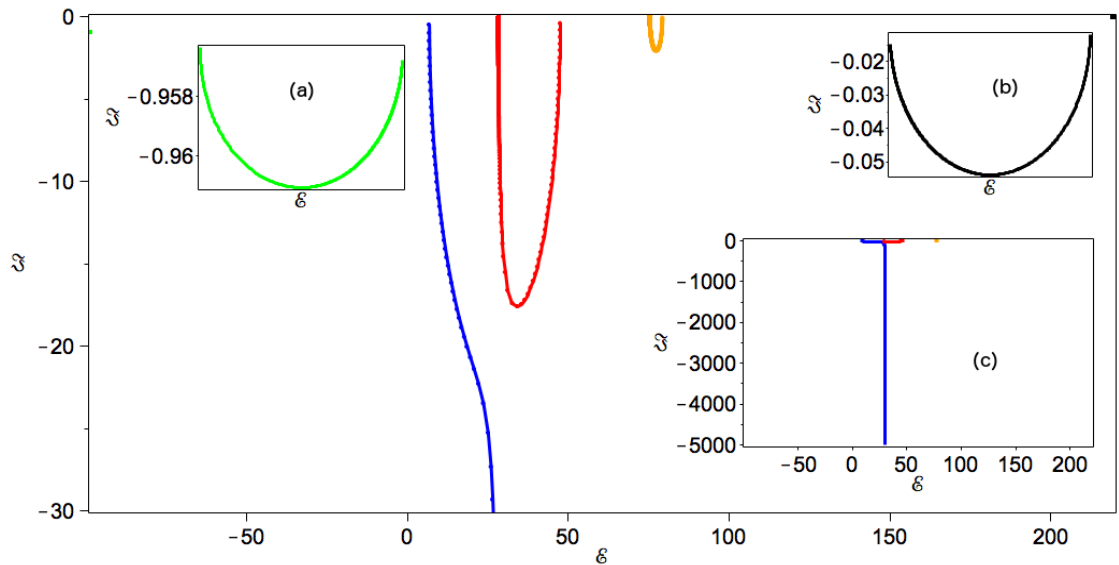}}
	\end{center}
	\caption{(Color online) The trajectories of five eigenvalues in the complex eigenenergy plane (${\mathfrak I}= \Im (\tilde E)$, ${\mathcal E}= \Re (\tilde E)$), as $\Gamma_4$ increases from 0 to 10000. The superradiant eigenstate (blue curve) is ``unstable" (its decay width, $|{\mathfrak I}|$, increases when $\Gamma_4$ increases). All other four eigenstates  become trapped. (Their decay widths are bounded above.) The insets, (a) and (b), show the detailed behavior of two eigenvalues located in the upper left and the upper right. The insert (c) demonstrates the behavior of the eigenvalues for large values of $\Gamma_4$. ($\varepsilon_0=-90,~\Gamma_0=2$).
		\label{G6}}
\end{figure*}

 When discussing the ST, the diagram of the trajectories of the eigenenergies in the complex plane (${\mathfrak I}= \Im (\tilde E)$, ${\mathcal E}= \Re (\tilde E)$) can produce additional insight to the features of the ST \cite{Zel1,Zel2}. We present this diagram in Fig. \ref{G6}, for five eigenenergies (indicated by different colors) parametrized by the ET rate, $\Gamma_4\in[0,10^4]$, and for the fixed value of $\Gamma_0=2$.  
  The superradiant eigenenergy is indicated by the blue curve. Its decay width increases when $\Gamma_4$ increases. The eigenenergies of four other eigenstates become trapped, as $\Gamma_4$ increases. Their decay widths are bounded, and experience a maximum as a function of $\Gamma_4$.  The superradiant eigenstate (blue curve) overlaps with the neighboring eigenstate, indicated by the red color,  at $\Gamma^*_4\approx 84\rm ps^{-1}$. All other neighboring resonances do not overlap.
  The detailed behaviors of the trajectories of two eigenenergies, located in the upper left and the upper right, are shown in  the insets, (a) and (b). The insert (c) demonstrates the behavior of the eigenenergies for large values of $\Gamma_4$. The following parameters were chosen: $\varepsilon_0 =-90\rm ps^{-1}\approx -59.4\rm meV$ and $\Gamma_0 =2\rm ps^{-1}$. 
 
 \subsection{Localization of the eigenfunctions}
 
\begin{figure*}[ht]
\begin{center}
\scalebox{0.45}{\includegraphics{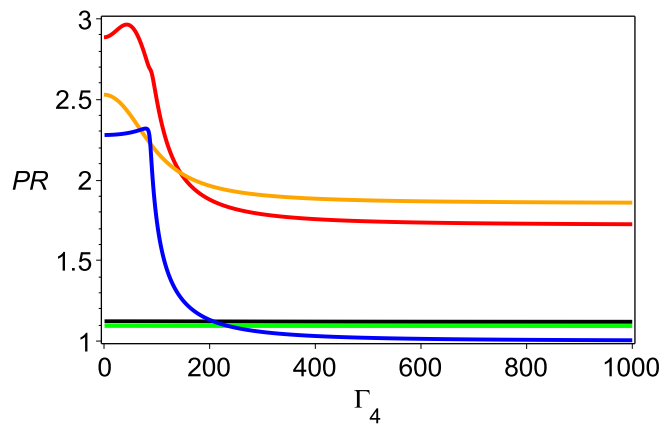}}
			\end{center}
		\caption{(Color online) The participation ratios  of the eigenstates of the effective non-Hermitian Hamiltonian, $\tilde {\mathcal H}$,  as functions of $\Gamma_4$. ($\varepsilon_0=-90,~\Gamma_0=2$).
		\label{PR}}
\end{figure*}

 To study numerically the localization properties of the eigenstates, we use the participation ratio (PR) \cite{IKM} of the eigenfunction, $|\psi_\alpha \rangle$, defined as, 
\begin{eqnarray}
PR_\alpha = \Big (\sum_n |\langle n| \psi_\alpha \rangle|^4 \Big )^{-1}.
\end{eqnarray}
Its value varies from 1, for a fully localized eigenstate, to 5, for a fully
delocalized eigenstate.  In Fig. \ref{PR}, the PRs for all five eigenfunctions are presented, as functions of $\gamma_4$. As one can see, the superradiance state (blue curve) is partly delocalized for relatively small values of $\Gamma_4<\Gamma_4^*\approx 84\rm ps^{-1}$, and it becomes localized at the site, $|4\rangle$, for large values of $\Gamma_4$. The sub-radiant eigenstates also experience reorganization in the vicinity of $\Gamma_4<\Gamma_4^*\approx 84\rm ps^{-1}$, but they remain partly delocalized for large values of $\Gamma_4$. These sub-radiant eigenstates contribute to the NPQ mechanism for large values of $\Gamma_4$. 
 
\begin{figure*}[ht]
\begin{center}
\scalebox{0.415}{\includegraphics{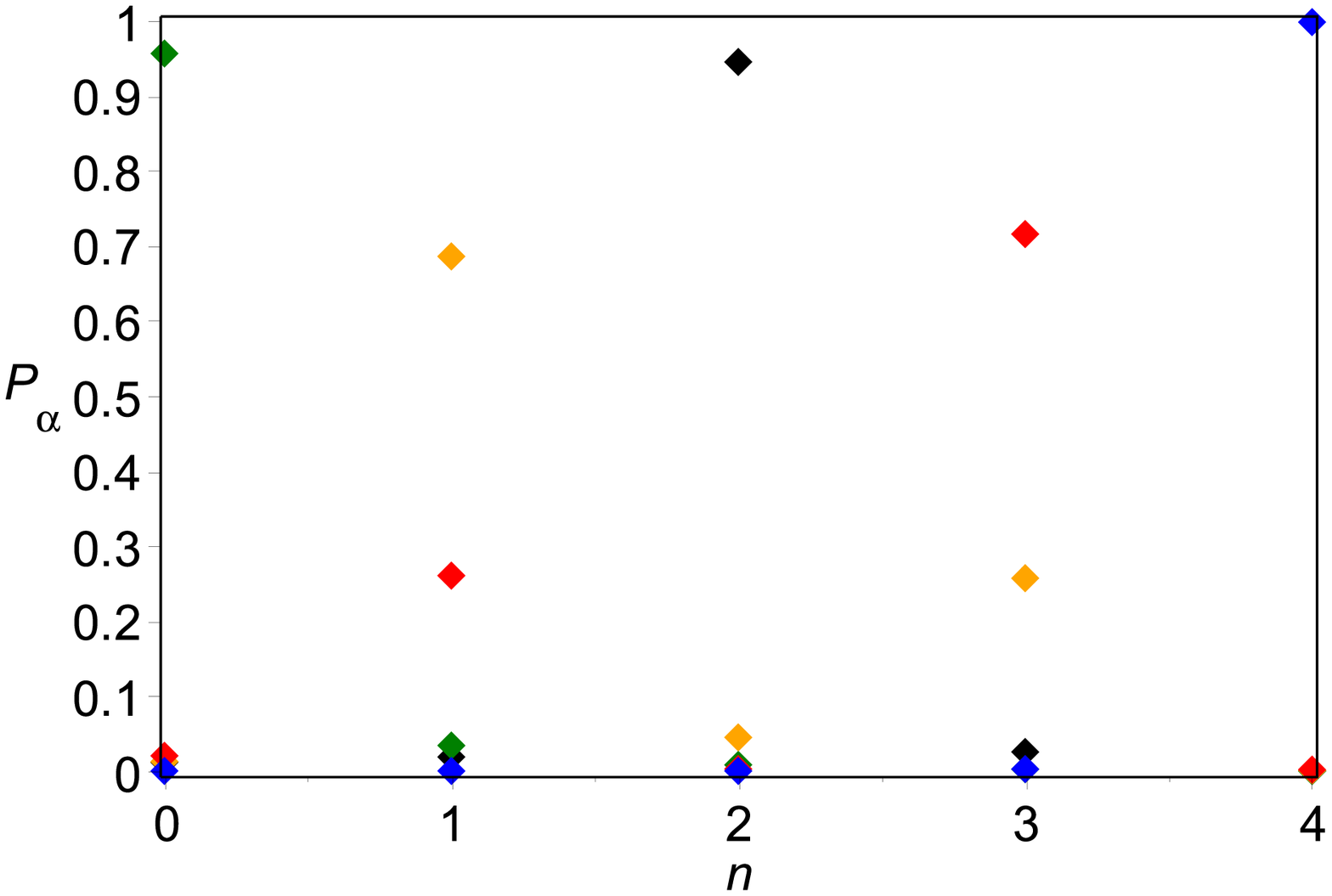}}
(a)
\scalebox{0.425}{\includegraphics{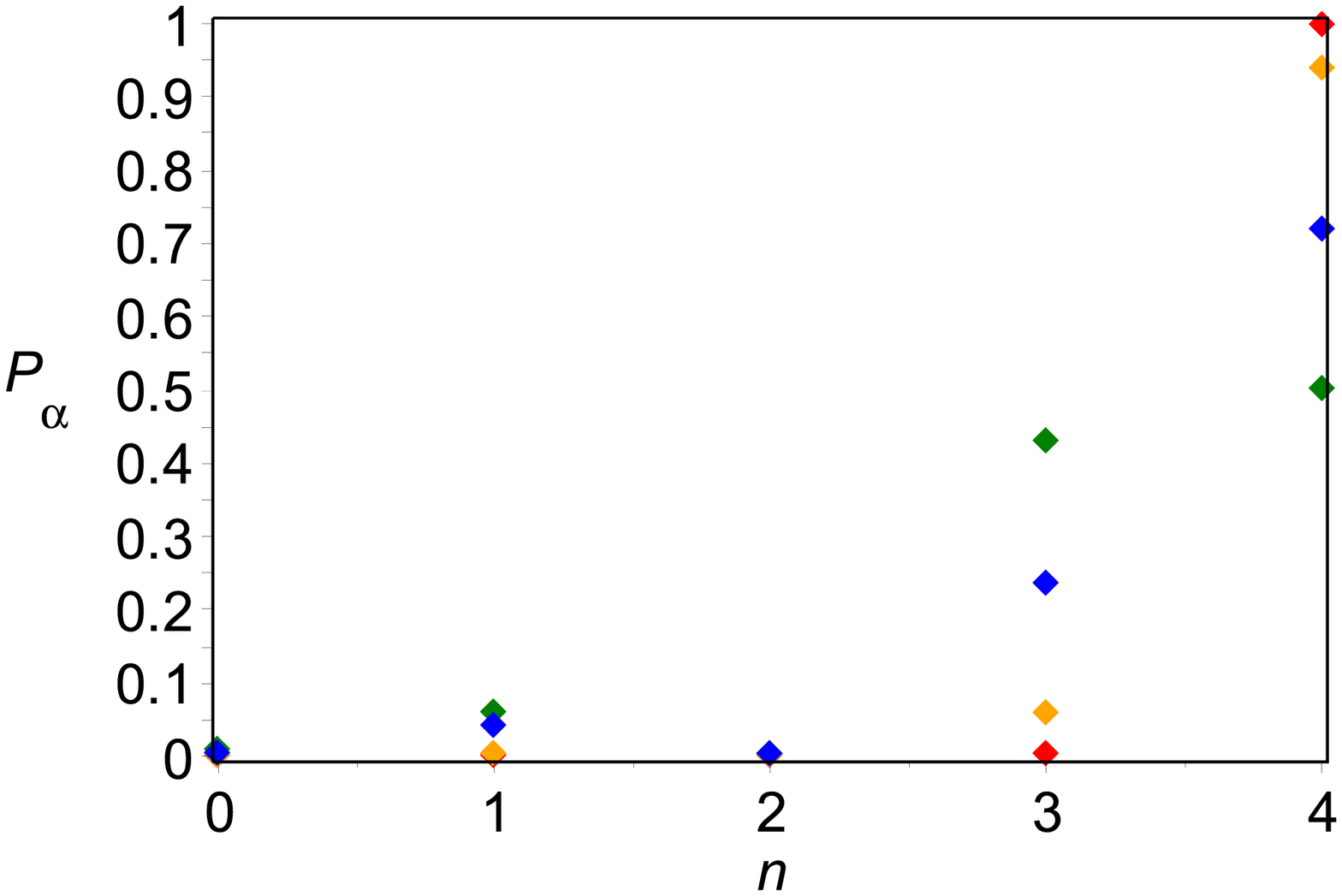}}
(b)
			\end{center}
		\caption{(Color online) (a)  Localization of the eigenstates ($\Gamma_4 = 10^3$).
		(b) Localization of the eigenstate with the largest decay width: $\Gamma_4=10$ (green), $\Gamma_4=100$ (blue), $\Gamma_4=200$ (orange),  $\Gamma_4=10^3$ (red). ($\varepsilon_0=-90,~\Gamma_0=2$). 
		\label{G5}}
\end{figure*}

 In Fig. \ref{G5}a, we demonstrate the localization properties of  eigenstates, for a moderate value of $\Gamma_0=2\rm ps^{-1}$, and for a relatively large value of $\Gamma_4=10^3\rm ps^{-1}$. The colors correspond to those used in Figs. \ref{G6} and \ref{PR}. On the vertical axis, the probability, $P_\alpha(n)=|\langle n|\psi_\alpha\rangle|^2$, is presented, of the population of the site for a given eigenfunction,  $|\psi_\alpha\rangle$. Different colors indicate different eigenfunctions. As one can see, the superradiant eigenstate, indicated by the blue curve, is localized mainly on the CTS, $|4\rangle$. This is the main reason why this eigenstate, which has the largest decay width, does not contribute significantly to the ST, or to the NPQ dynamics. (See green curve in Fig. \ref{A3}b.) Indeed, the initial condition corresponds to the population of the $Chla_5^*$ (site $|1\rangle$). Then, at large values of the ET rate, $\Gamma_4$, this initially populated state, $|1\rangle$, only weakly overlaps with the superradiant eigenstate.  So, during the finite time, $\tau=5\rm ps$ (as in Figs. \ref{A3}a and \ref{A3}b)), the efficiency of the NPQ, $\eta_4(\tau=5\rm ps)$ is relatively small. The NPQ efficiency remains relatively small, for large values of $\Gamma_4$, in the presence of noise. (See,  Figs. \ref{A3}a and \ref{A3}b.) 
 
 In Fig. \ref{G5}b, we show the localization properties of the superradiant eigenstate (with the largest decay width), for $\Gamma_2=2\rm ps^{-1}$, and for four values of the ET rates, $\Gamma_4\rm$. As one can see, for relatively the small value, $\Gamma_4=10\rm ps^{-1}$ (green color), the superradiant eigenstate is inhomogeneously delocalized over all five sites. The value $\Gamma_4=100\rm ps^{-1}$ (blue color) corresponds to the vicinity of the ST, and to the best performance of the NPQ mechanism. The superradiant eigenstate is still inhomogeneously delocalized over all five sites. When $\Gamma_4$ increases ($\Gamma_4=200\rm ps^{-1}$, orange color), the localization of the superradiant eigenstate on the CTS, $|4\rangle$, increases. For large values of $\Gamma_4=10^3\rm ps^{-1}$ (red color), the superradiant eigenstate is mainly localized on the CTS (site $|4\rangle$).

\section {Conclusion}

Our main result is the demonstration of the relation between the two well-known phenomena: the nonphotochemical quenching (NPQ) mechanism in photosynthetic organisms and the superradiance transition (ST), that occurs in many physical and biological systems. For concreteness, we analyzed this relation by using a relatively simple model for the NPQ by the charge transfer state (CTS), in the CP29 minor LHC of the PSII. The NPQ protocol we used, is mainly based on the experimental results discussed in \cite{Ahn}. 

In the absence of noise, our  model is described by an effective non-Hermitian Hamiltonian, which is represented by a non-Hermitian $5\times5$ matrix.  The eigenenergies of this matrix are complex numbers. According to the accepted terminology, their real parts represent the positions of the resonances (poles), in the complex energy plane. The absolute values of their imaginary parts are the decay widths of the resonances. The ST occurs when these resonances begin to interact (overlap). In this case, the superradiant eigenstates accumulate large decay widths. So, these superradiant states contribute coherently to the radiation into the continuum spectra (sinks). All other eigenstates (sub-radiant) have small decay widths, and are long-lived. However, for some range of parameters, the superradiant eigenstates  become localized at the sites which do not overlap with those sites which are initially populated. In this case, the superradiant eigenstates practically do not contribute to the ST (in spite of the fact that these eigenstates have the largest decay widths). This competition between the decay widths and the delocalization of the superradiant eigenstates is well-know in the field of the SR. 
Since our model includes two sinks, two STs  occur for different values of parameters. One ST occurs in the damaging channel, and the other occurs in the charge transfer channel. This latter one contributes to the NPQ mechanism. 

The efficiency of the NPQ by the CTS mainly depends on two factors: the ET rate to the sink connected to the site of the CTS, and on the probability accumulated at the excitonic site, where the CTS is localized. When the ET rate to the sink associated with the CTS is small, the NPQ efficiency is also small. However, when this ET rate is very large, the superradiant eigenstate becomes strongly localized at the excitonic CTS, and it practically does not overlap with the initially populated chlorophyll dimer. Then, the NPQ efficiency is also suppressed in this case. Only for the intermediate values of the ET rate in the sink associated with the CTS, does the NPQ efficiency approaches its maximum. 

We demonstrated that the maximum of the NPQ efficiency is caused by the ST to the sink associated with the CTS. However, experimental verifications are required in order to determine whether or not the NPQ regime is associated with the ST transition for real photosynthetic complexes. Indeed, it can happen that, in the photosynthetic apparatus, the NPQ regime occurs in the ``non-optimal" region of parameters, and it could be independent of the ST.  

The results obtained in this paper can easily be generalized to more complicated networks of interacting chlorophyll and carotenoid light-sensitive molecules, and for more complicated random protein environments.  The experimental verification of the relation between the NPQ and the ST phenomena  would be of significant interest for both better understanding of the NPQ mechanisms and for bio- and nano-engineering.

\begin{acknowledgments}
This work was carried out under the  auspices of the National Nuclear Security Administration of the U.S. Department of Energy at Los Alamos National Laboratory under Contract No. DE-AC52-06NA25396. A.I.N. and G.L.V. acknowledge the support from the CONACyT. G.P.B. and R.T.S. acknowledge the support from the LDRD program at LANL.
\end{acknowledgments}

\end{article}

\end{document}